%% file: m914.tex
\documentclass{ecai} 
\usepackage{latexsym}
\usepackage{amssymb}
\usepackage{amsmath}
\usepackage{amsthm}
\usepackage{booktabs}
\usepackage{enumitem}
\usepackage{graphicx}
\usepackage{color}
\usepackage{algorithm}
\usepackage[switch]{lineno}
\usepackage{multirow}
\usepackage[table]{xcolor}
\usepackage{float}
\usepackage{booktabs,caption}
\usepackage[flushleft]{threeparttable}
\usepackage{tikz-cd}
\usepackage{xcolor}
\usepackage{tcolorbox}
\usepackage{stfloats}
\usepackage{algpseudocode}

\newcommand{\BibTeX}{B\kern-.05em{\sc i\kern-.025em b}\kern-.08em\TeX}

\begin{document}

%%%%%%%%%%%%%%%%%%%%%%%%%%%%%%%%%%%%%%%%%%%%%%%%%%%%%%%%%%%%%%%%%%%%%%%%

\begin{frontmatter}

%%% Use this command to specify your submission number.
%%% In doubleblind mode, it will be printed on the first page.

\paperid{123} 

%%% Use this command to specify the title of your paper.

\title{Big-Thick Data generation via reference and personal context unification}
 
\author[A]{\fnms{Fausto }~\snm{Giunchiglia}}

\author[A]{\fnms{Xiaoyue}~\snm{Li}\thanks{Corresponding Author.}}

\address[A]{Information Engineering and Computer Science, University of Trento, Italy}

\address[]{\{fausto.giunchiglia, xiaoyue.li\}@unitn.it}

%%% Use this environment to include an abstract of your paper.

\begin{abstract}
Smart devices generate vast amounts of \textit{big data}, mainly in the form of sensor data. While allowing for the prediction of many aspects of human behaviour (e.g., physical activities, transportation modes), this data has a major limitation in that it is not \textit{thick}, that is, it does not carry information about the context within which it was generated.  \textit{Context} – what was accomplished by a user,
how and why, and in which overall situation – all these factors must be explicitly represented for the data to be self-explanatory and meaningful.
In this paper, we introduce \textit{Big-Thick Data} as highly contextualized data encoding, for each and every user, both her \textit{subjective} personal view of the world and the \textit{objective} view of an all-observing third party taken as reference. 
We model big-thick data by enforcing the distinction between \textit{personal context} and \textit{reference context}. 
We show that these two types of context can be \textit{unified} in many different ways, thus allowing for different types of questions about the users' behaviour and the world around them and, also, for multiple different answers to the same question. 
We validate the model with a case study that integrates the personal big-thick data of one hundred and fifty-eight University students over a period of four weeks with the reference context built using the data provided by OpenStreetMap. 
\end{abstract}

\end{frontmatter}

%%%%%%%%%%%%%%%%%%%%%%%%%%%%%%%%%%%%%%%%%%%%%%%%%%%%%%%%%%%%%%%%%%%%%%%%

\section{Introduction}
\input{section/1-Introduction}

\section{Related Work}
\label{sec:Related Work}
\input{section/2-RelatedWork}

\section{Reference and Personal Context}
\label{sec:Defining Context}
\input{section/3-DefiningContext}

\section{Context Unification}
\label{sec:4-unification}
\input{section/4-unification}

\section{Context Observation}
\label{sec:5-observ}
\input{section/5-observation}

\section{Case Study}
\label{sec:Case Study}
\input{section/6-CaseStudy}

\section{Conclusion}
\label{sec:Conclusion}
\input{section/7-Conclusion}

\newpage
\begin{ack}
The first and most important thanks goes to Matteo Busso for inventing the term \textit{big-thick data} and for producing the first version of the picture in Figure \ref{bigthick}, reported here essentially unmodified. The work described in this paper could be done thanks to the amazing group of colleagues in the KnowDive Group: Mayukh Bagchi, Simone Bocca, Andrea Bontempelli, Ali Hamza, Leonardo Havier Malcotti, Ivan Kayongo, Alessio Zamboni and Haonan Zhao.
This research has received funding from the European Union’s Horizon 2020 FET Proactive project “WeNet – The Internet of us”, grant
agreement No 823783.
\end{ack}
\bibliography{mybibfile}

\end{document}

%% file: section/1-Introduction.tex
Smart devices, e.g., smartphones or smartwatches, allow for the collection of a wide set of large-scale sensor data, e.g., GPS, Bluetooth, WIFI or accelerometer. This type of data, often referred to as (a specific kind of) \textit{big data} \cite{bornakke2018big}, has been widely exploited, for instance, in Human Activity Recognition \cite{sousa2019human}, Health Monitoring \cite{sheikh2021wearable} and Autonomous Vehicles \cite{yeong2021sensor}. However, this type of data is often used ‘out of context’ and this substantially obscures its meaning and, therefore, diminishes its value \cite{boyd2012critical}, in particular when trying to understand human behaviour, e.g., one's social or personal life, which is always context-sensitive. \textit{Context} – what was accomplished by a person,
how and why, and in which overall situation – all these factors must be explicitly represented for the data to be self-explanatory and meaningful \cite{giunchiglia2017personal}. In particular, these factors become necessary if one wants to use the same dataset for multiple predictions, where the same sensor value may stand for two completely different contextual situations. So, for instance, a professor and a student may be in the same location, e.g., the university, with different purposes, the first for work, the second for study or because looking for a friend, while the former was in the same location during the last week-end because she wanted to collect her tennis racket which she left there on Friday. Meaningful, lifelong human-in-the-loop, human-machine interactions need this level of information richness.

To address the problem of data de-contextualization, we turn to the notion of \textit{Big-Thick Data}. Big-thick data is big data complemented with \textit{thick data}, that is, \textit{observational data about context which allow to reflect upon how and why people do what they do}. We build \textit{Observation Contexts}, as we call them,  to represent big-thick data based on two main components, one or more users' \textit{Personal Contexts} and a \textit{Reference Context}.
A \textit{personal context}, one for each and every user, encodes the user's \textit{subjective} view of the world, e.g., where she is, what she is doing, why she does it, who she is with, her mood \cite{giunchiglia2017personal}. Personal contexts are different for different users, also when involved in the same activity, and are also different for the same user at different times, this because of their evolving activities. We model personal contexts, in time, as \textit{Personal Big-Thick Data} obtained by integrating \textit{Personal Big-Data}, e.g., sensor data or data from social media, with \textit{user-provided descriptions} of the current situation, for instance, in terms of crowd-sensing \cite{abualsaud2018survey}, human answers to machine questions \cite{,giunchiglia2018mobile}, people's self-reports \cite{KD-2021-Zhang-putting}, or information from the phone's personal contacts or agenda \cite{zhao2024human}.  

A \textit{reference context} provides a user-independent
 \textit{objective} all-encompassing view of a third-party observer. It keeps track of the environment within which users are operating, defined in terms of a \textit{reference observation period}  and a \textit{reference location}. Examples of reference observation periods are one day, one week or one year. Examples of reference locations are home, the city of Trento, or Italy. 
Which is the ‘right' one depends on the \textit{purpose}. For instance, the reference location could be home if the user is watching the television, the street outside if she is at the window, or Trento if she is driving to the university. Each location determines events and entities, each person entity with its own personal big-thick data. The reference context can be built out of any type of spatio-temporal (big) data, e.g., coordinates, images, labels, as from, e.g.,  OpenStreetMap (OSM)\footnote{\url{https://www.openstreetmap.org}.} or the Italian Spatial Data portal\footnote{\url{https://www.agid.gov.it/en/data/spatial-data}.}.

The \textit{observation context} is built out of (a part of) the reference context and (parts of) one or more personal contexts based on shared \textit{identifying information}, e.g., names, identifiers, spatio-temporal coordinates. %The reference context is the key pivotal element in this process, what allows to ‘absorb' subjectivity into objectivity. 
The idea is to compose the subjective information of personal contexts into the objective perspective of the reference context. We call this process, \textit{context unification}. We implement context unification as a flexible process, which is \textit{configured} as a function of the specific \textit{purpose}, as defined in \cite{giunchiglia2022popularity}. Some examples of purpose are, for instance, the need of answering a specific query or the need of learning about the behaviour of a certain class of people, where the observation context may be tuned to a specific person or a group of people or to everybody we know is inside the reference location.

The main contributions of this paper are as follows:

\vspace{-0.2cm}
\begin{enumerate}
    \item The notion of \textit{big-thick data} and its operationalization in terms of \textit{observation, }\textit{personal} and \textit{reference context};
    \item A methodology for the \textit{purpose-driven} generation the observation context via \textit{context unification};
    \item A methodology, that we call \textit{context observation}, for exploiting the observation context with multiple different purposes.
\end{enumerate}
\vspace{-0.2cm}
\noindent
We perform a first assessment of the approach proposed via a case study where we unify a dataset, called SmartUnitn2 (SU2),\footnote{A description of the project which generated the SU2 dataset can be found at the link \url{https://datascientia.disi.unitn.it/projects/su2/}. This page provides also information about how to download it. \label{foot:SU2}} describing the behaviour of a large sample of university students, with a dataset generated from OSM.\footnote{A description of the work described in this paper can be found at the link \url{https://datascientia.disi.unitn.it/projects/su2osm/}. This page provides a link to the OSM and SU2 datasets used in the case study.}
The structure of this paper is as follows. Section \ref{sec:Related Work} describes the related work. Section \ref{sec:Defining Context} introduces reference context and personal context. Section \ref{sec:4-unification} explains context unification. Section \ref{sec:5-observ} describes context observation. Section \ref{sec:Case Study} provides the case study. Section \ref{sec:Conclusion} concludes the paper.

%% file: section/2-RelatedWork.tex
We organize the section into \textit{big-thick data} and \textit{context}.

\vspace{0.1cm}
\noindent \textbf{Big-Thick Data.}
The notion of \textit{personal big data} came up with the explosion of digital records extracted from smart devices, mainly via sensors. 
While it facilitates predictive analytics that far exceed human cognitive capabilities, big data does not support human observation and interpretation in context \cite{ang2019integrating}. 
Big data is always too poor in its contextualization to describe many interesting aspects of people behaviour, role and motivation for action \cite{Blank2008Online}. This is why Bornakke and Due, in \cite{bornakke2018big}, talk of big data as \textit{big thin Data}.
% Uses of thick data
Differently from big data, \textit{thick data} is constructed from observations of the context in which human behavior occurs, historically, mainly in the form of people interviews and extensive self-reports \cite{bornakke2018big}. The idea of thick data originated from what in Anthropology are called \textit{thick descriptions} of the world \cite{geertz2008thick}, that is, ethnographically
collected and analysed observational, contextually rich, \textit{small data}. Despite the small size, thick data is still rich enough in content to enable researchers to understand and reflect upon the scenario within which people act and behave. A major problem remains which relates to the cost of generation of ethnographically generated thick data.

\begin{figure}[htp!]
\vspace{-0.2cm}
\centering
\includegraphics[scale=0.60]{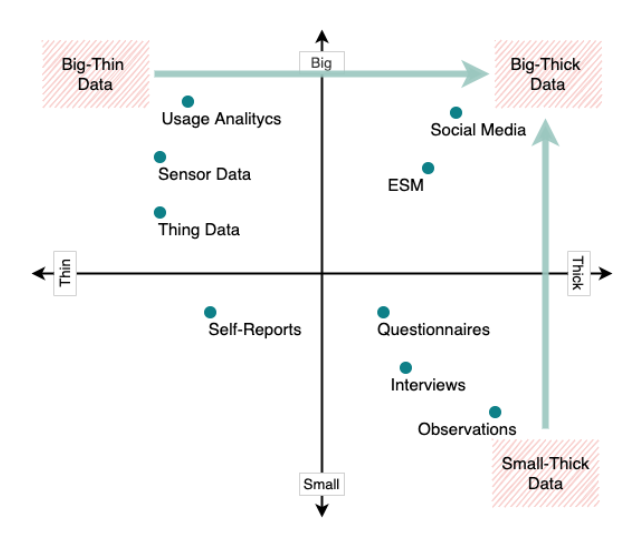}
\caption{ \centering Big-Thick data.}
\vspace{0.4cm}
\label{bigthick}
\end{figure}

% big-thick data
Recent work has suggested the integrated usage of big computational quantitative data and small embodied qualitative thick data \cite{blok2014complementary,boellstorff2013making}, with the recent introduction of the concept of \textit{Big-Thick Blending} \cite{bornakke2018big}. However, this work is mainly qualitative. The notion of \textit{big-thick data}, as introduced in this paper, together its computational realization in term of reference and personal context, extends and operationalizes the idea of big thick blending. Figure \ref{bigthick}, which is an evolution of Figure 2 in \cite{bornakke2018big}, defines big-thick data as the convergence of Big-Thin Data, e.g., Usage Analytics, Sensor data, general Internet-of-Things (IoT) data (Thing Data), Small-Thin Data, e.g., Self-Reports, Small-Thick Data, e.g., Observations, Interviews and Questionnaires, and some first examples of Big-thick data, e.g., Social Media and Ecological-Momentary-Assessment / Experience-Sampling-Method (EMA/ESM) Data.

\noindent
Of course this is just the beginning and there are still important open problems, in particular when one focuses on lifelong, human-in-the-loop human-machine interactions. A crucial issue is how to decrease the cost while increasing the quality of big-thick data collection. %Related to this problem is the largely sub-optimal user behaviour. 
For example, users often do not read, or do not answer, or provide wrong answers to machine-asked questions \cite{bison2024impacts}, or turn-off their data collection APP. 
\cite{KD-2022-Bontempelli-lifelong} provides a general overview of how we approach the problem.
\cite{bison2024impacts} describes first results towards the management of long response times. \cite{KD-2020-Bontempelli1} deals with the problem of mislabelling
while \cite{zhao2024human} describes an early version of an APP which supports the user in providing meaningful answers.

\vspace{0.1cm}
\noindent
\textbf{Context.} The notion of context has a long history, and has been studied extensively in multiple research areas. The area where it was first introduced is, as far as we know, Knowledge Representation (KR), starting from  John McCarthy's Turing Award lecture \cite{mccarthy1987generality}. Here context was proposed, together with non-monotonic reasoning, as a key element for the formalization of commonsense reasoning. In \cite{giunchiglia1993contextual} context and multi-context systems (MCSs) were introduced as personal representations of real world situations, as defined by John McCarthy \cite{mccarthy1981some}. The idea of multiple subjective views of the same objective reality was first introduced in \cite{giunchiglia1998local} using the example of the \textit{magic box}, later recollected in \cite{brewka2007equilibria}. In \cite{giunchiglia1998local} a \textit{view} was defined as a partial \textit{representation} of the world consisting of as a set of facts describing the user current perspective.  
Later on, Brewka et al. \cite{brewka2018reactive} used MCSs for the representation of multiple heterogeneous knowledge sources, with information flows
allowing for reasoning across multiple contexts. The notion of reference context relates to Guarino and Guizzardi's notion of \textit{scene}  \cite{guarino2016relationships}. The key difference, using the terminology in \cite{guarino2016relationships}, is that a scene is a perdurant while the reference context is an endurant. The connection is in the fact that a scene can be seen as the perdurant of the reference context.

In IoT research, sensor data is exploited to enable \textit{context recognition} from sensor time-series data, where context is then used to learn about the user behaviour. In this work, the possible context dimensions include locations, activities, body posture, and more \cite{vaizman2018context}.
For example, accelerometer data are used to detect physical activities, e.g., walking or running \cite{bettini2020caviar,shen2023transfer,vaizman2018context}. 
A Multi-Layer Perceptron (MLP) that uses multi-modal sensors (smartphone accelerometer, smartwatch accelerometer, phone gyroscope, phone audio, etc.) is described in \cite{vaizman2018context} which allows to simultaneously predict many diverse context labels, e.g.,  people's body-states, home activities and environment.
At the boundary between IoT and KR, Giunchiglia et al. have used context to model personal data streams \cite{giunchiglia2017personal,%giunchiglia2018personal,%
KD-2022-Giunchiglia.a}. 
The work in this paper builds upon the previous work in KR and IOT,  and in particular \cite{giunchiglia2017personal,%giunchiglia2018personal,%
KD-2022-Giunchiglia.a}, and it constitutes an attempt towards the generation of big-thick data,  up to the quality which is needed in order to support meaningful human-in-the-loop human-machine interactions.

The notion of context has also been extensively studied by HCI community as ‘any information that can be used to characterise the situation of an entity' \cite{abowd1999towards,dey2001understanding}. The underlying intuition was to use context in order to facilitate richer and easier human-machine interactions \cite{dey2001conceptual}. For example, the Activity River \cite{aseniero2020activity}, a personal visualization tool, enables people to visualize historical and contextual data (e.g., activities), flexible planning and logging, etc. The relevance of this work stems from the fact that human-in-the-loop human-machine interactions need the high-quality interfaces and interactions studied and developed by this community.

%% file: section/3-DefiningContext.tex
Let us think of the \textit{world} as an infinite set of, continuously evolving, three-dimensional spatial regions, that we call \textit{locations}, 
and, inside each location, as an infinite set of, continuously evolving, mono-dimensional temporal regions that we call \textit{events}. Events have a \textit{duration}  defined by a \textit{start-time} and an \textit{end-time}.
Then, let us take the world as being populated by \textit{entities}, e.g., people, trees, homes, cities, streets, anything which we can think as having a spatial and a temporal extension. We follow \cite{KD-2017-ICCMF} and we think of entities as being identified by two key components, each with its own properties, that is
 \textit{objects}, which define the spatial regions occupied by entities, and
\textit{functions}, which define a specific set of entity \textit{expected actions}, with actions being the mechanism by which events change the world. Thus for instance, a \textit{car} and a \textit{bus} are two entities associated to two different objects both performing the same function of a \textit{vehicle}, which is characterized by the action of \textit{carrying} people around.  
Dually, we can also think of (the body of) a person as an object supporting many functions and corresponding entities and actions, e.g., a student reading a textbook or a driver driving a taxi.
Finally, taking an example from \cite{KD-2017-ICCMF}, an entity implementing the function of a \textit{chicken coop} may consist of as a little wooden house or of the body of an old car. In this latter case the same object was first associated to the entity car and the to entity chicken coop. 
The key observation is that there is a many-to-many relation between objects and functions, and any such combination defines an entity, not necessarily the same.

We say that, in an object is \textit{partIn}, or \textit{populates}, a location if it is inside the region identifying the location. Similarly,  in a 
given moment in time, an entity is \textit{partIn}, or \textit{populates}, an event if the entity is inside the spatial region of the location of the event and within its temporal region. 
Entities are associated, for each location and event they populate with, respectively, \textit{(spatial) coordinates} and \textit{(temporal) coordinates}. GPS coordinates and the local time of a time zone, while not being the only possible choice, provide the coordinates for any possible triple <event, location, entity>. Notice that there is no need for locations or events to be positioned with respect to some external spatio-temporal coordinate system. The positioning is only of entities inside locations and events. This captures the intuition that, for instance, when you are at home, what you do depends only on the entities inside home, e.g, the television, and on their evolving state, e.g., the television being turned on, and not, e,g, on the location of the apartment in the city. And similarly for time.

People are entities which have an internal \textit{representation} of the world and use it to reason about it and take action. Following \cite{giunchiglia1993contextual}, we assume that this mental representation is organized in \textit{contexts} where (quote) \textit{`... a context is a theory of the world which encodes an individual's subjective perspective about it'}. According to \cite{giunchiglia1993contextual}, contexts are not \textit{situations}, where, following \cite{mccarthy1981some}, (quote) \textit{`... a situation $s$ is the complete state of the universe at an instant of time'}. In other words, people have partial, possibly incorrect, views of the world while situations establish what is the case, thus providing a single reference point for comparing the contents of contexts. However, situations are not accessible, we can only build mental representations about them. In the following we call \textit{personal contexts} all those contexts, carrying a subjective representation of the world, which satisfy the definition of context from \cite{giunchiglia1993contextual}, while we call \textit{reference contexts} all those contexts carrying an \textit{objective} representation of situations, as defined in \cite{mccarthy1981some}. Notice that here we talk about subjectivity and objectivity in a somewhat limited form, with reference to the fact that people have \textit{partial knowledge} of the world, thus leaving out issues related to the subjectivity of, e.g., opinions or sentiments. Thus, we say that we have \textit{objective knowledge} if everybody in the target audience knows or has the means for knowing about it. Dually, \textit{subjective knowledge} is known only by a few specific subjects while the others may or may not know about it, in the latter case, being able to get to know about it only if told by those who know about it. Thus for instance, my home address is personal knowledge, while the name of the street where I live is objective knowledge.

In the following,  we
first introduce the reference context $C_R$ (Section \ref{sec:Reference Context Definition}) and then the  personal context $C_P$ (Section \ref{sec:Personal Context Definition}).
 
%%%%%%%%% 3.1 RC %%%%%%%%%%%
\subsection{The Reference Context}
\label{sec:Reference Context Definition}

We wrote above that reference contexts carry `... \textit{an objective} representation of situations' and \textit{not} that they carry
`...  \textit{the objective} representation of situations'.
This is because it is impossible to build a complete description of reality. Two objective representations of the same situation may differ in many dimensions, for instance, the level of detail, the level of partiality, the view point, the entities being considered, and so on \cite{KD-2017-ICCMF}. 
We follow an approach where we build the reference context based on a specific \textit{purpose}  where, following 
\cite{giunchiglia2022popularity}, loosely speaking, the purpose is connected to the specific target use, e.g. answering a specific query or recognizing a specific action. The process is similar to when one has the need of asking multiple questions, for instance: ‘what are the people involved in the current event  doing now and what will they be doing tomorrow, when in the same location?', ‘does the current location usually host similar events?', and, as a result,  she focuses on different fragments of her knowledge.

We proceed as follows. 
We start by choosing a \textit{Reference Spatial Region} $S$, defined as the set of points $(x,y,z)$ located inside the \textit{boundary} of $S$, where the boundary defines the \textit{inside / outside} $S$, with reference to a bigger location that is not considered because not purpose-relevant. Let us assume that $S$ contains a set of spatial regions, perceived as \textit{objects}, $O_1, ..., O_n$,
themselves defined by a boundary, an inside and an outside \cite{KD-2017-ICCMF}, that is
\begin{equation}
\vspace{-0.2cm}
O_i \subseteq S \ \ \ \ \ \text{with} \ \ \ \ \  1 \leq i \leq n,
\label{eq:L}
\end{equation} 

\noindent
Let us concentrate on a \textit{Reference Observation Period} $\Delta T_R$ where $t\in \Delta T_R$ measures how change happens within $S$. Then we define a \textit{Spatio-temporal Context} $C_S$ as:
\vspace{-0.1cm}
\begin{equation}
C_S(t)= \langle S(t), \{O(t)\}_S \rangle\ \ \ \text{with}\ \ \  t\in \Delta T_R
\label{eq:C_R1}
\vspace{-0.1cm}
\end{equation}
\noindent 
where $\{O(t)\}_S$ is the set of objects $O_i$ satisfying Eq. (\ref{eq:L}). Intuitively, Eq. (\ref{eq:C_R1}) tells us that $C_S$ consists of a set of objects located within $S$, and that both $S$ and the objects $O_i$'s change
during $\Delta T_R$.
Time variance is a fact of life. Everything continuously changes. This is a major source of subjectivity as the same object looks different at different times. We deal with this problem by requiring that, during $\Delta T_R$ we have \textit{Time Invariance.}  That is,  $\Delta T_R$ must be such that the \textit{Reference Location} $L_R$, that is, the location associated to the spatial object $S$, and the selected objects in $L_R$ do not change, that is, they keep the same \textit{selected spatial properties}, e.g., position, shape and color. Period, objects and spatial properties are selected as a function of the purpose.
We move from $C_S$ to the \textit{Reference Context} $C_R$ as follows. Let $e_i^j$, with $j=1, ...,m$ be the $j$-th entity associated to the object $O_i$. Then we have:

\vspace{-0.2cm}
\begin{equation}
C_R  =   f(C_S(t))   =  \langle L_R ,  \{e\}_R \rangle\ \ \ \text{with}\ \ \  t\in \Delta T_R
\label{eq:C_R2}
\end{equation}

\noindent
where: $C_R$ is time-invariant, $\{e\}_R\subseteq \{ e(O)\}_S$ is the set of entities of $C_R$, with $\{ e(O)\}_S$ the set of the entities associated to the objects of $C_S$,
 and $f$ is a projection function from $C_S$ to $C_R$ enforcing \textit{objectivity}. The definition of $f$ is up to the modeler, with the proviso that it must satisfy the following set of constraints.

\vspace{-0.2cm}

\begin{description}

\item \textit{From objects to entities.} $L_R$ and $e\in \{e\}_R$
should be chosen to fit the purpose. For instance the region $S$ associated to the Trento spatial region can be thought of as the location represented by a geographical map, or by an administrative map.
%; or the empty space of a room can be associated to a bedroom or a storage room. 
Similarly the object corresponding to (the body of) a person can be thought, e.g., as a father or as a professor;

\item \textit{Completeness wrt. the users'.} $C_R$ should provide enough detail to ground the different subjective views of all the users. Users should be able to determine whether their representation is consistent with that of the reference context and that of any other user;

\item\textit{Localization.} $C_R$ should describe the smallest possible location satisfying the previous properties. 
\end{description}
\vspace{-0.2cm}
\noindent
Some observations. The first is about how \textit{objectivity}  is enforced. 
\textit{Time invariance} allows to generate shared knowledge which does not change in time. This is a very robust form of objective knowledge, the easiest to manage and scale. 
This may seem a strong requirement, but notice that most spatial entities, e.g., streets, cities, monuments, the furniture in an apartment, change very rarely. Plus, this approach can be extended to manage what one could call \textit{objective events}, that is events everybody knows about, for instance a concert which has been organized long before its occurrence, and extensively advertised so that everybody knows about it. The idea is to decompose the duration of the observation period $\Delta T_R$ into a sequence of smaller periods, each corresponding to a time independent $C_R$, via a time-aware versioning mechanism. %This approach is scalable and easy to automate.
Furthermore, and this is the second key element towards the enforcement of objectivity, when moving \textit{from objects to entities}, only the purpose-relevant functions and actions are selected.
Thus, for instance, for the city of Trento we may have multiple reference contexts, one providing information about moving around, one about health related of university related facilities, one about points of interest, and so on \cite{KD-2022-Bocca}. 
This means, furthermore, allowing for the computation of the relevant subset of spatial relations among the objects in $L_S$. Some examples are: positioning (e.g., via coordinates),  relative positioning (e.g., \textit{Right} or \textit{Above}), proximity, reachability, color, or shape. 
The request of \textit{Completeness wrt. the users'} allows to enforce a general mechanism for the comparison of the user contexts. It allows the reference context to take the role of an oracle capable of deciding what is true and what is false among the facts stated inside personal contexts. 
\textit{Localization} is a key requirement for the reference context to work in practice. Thus, for instance, if the focus is what is happening at home, then the reference context should not include entities which are outside home. 

\subsection{The Personal Context}
\label{sec:Personal Context Definition}
People are some among the entities inside $L_S$, each of them with their own unique subjective view of the world, that we formalize as their own personal context $C_P$. Compared to $C_R$, $C_P$ has a few distinguishing features, as follows. Let $me$ be a generic person.

\begin{description}
 \vspace{-0.1cm}
 \item $C_P = C_P(me, C_R)$. $C_P$ depends on both $me$ and $C_R$.  Different $C_R$'s may generate different $C_P$'s even for the same $me$ and $S$;

    \item \textit{From inside to outside.} While $C_R$ describes the entities which are \textit{inside} $S$, $C_P$ describes the entities which are \textit{outside} the object of $me$, still \textit{inside} $S$;

    \item \textit{From no change to change.} Differently from $C_R$, $C_P$ considers also entities which change in time. 
    
\vspace{-0.2cm}
\end{description}
\noindent
We define the \textit{personal context} $C_P$ of $me$,given $C_R$, as follows:
\vspace{-0.1cm}
\begin{equation}
C_P = \langle C_R, \{E(L_R, t)\} \rangle \ \ \ \text{with}\ \ \  t\in \Delta T_R
\label{eq:C_RE1}
\vspace{-0.2cm}
\end{equation}
\noindent 
where $E(L_R,t)$ is a time-varying \textit{event} \textit{as perceived by $me$}, involving $me$, and occurring inside $L_R$, with $\{ E(L_R,t)\}$ being a set of such events. 
Any two events may occur in sequence or in parallel. These events model how situations, as subjectively perceived by $me$, evolve in time. 
Notice how Eq. (\ref{eq:C_RE1}) is the same as Eq. (\ref{eq:C_R2}) when one substitutes a set of unchanging entities with a set of continuously changing events.
This is the formalization, and generalization to a time-variant real world setting, of the idea of \textit{subjectivity as view point} of the \textit{magic box} \cite{giunchiglia1998local}. That is, there is an unchanging fully known objective reality, modeled by $C_R$, and multiple time-varying partial perspectives of this reality, each modeled by the set of events in which $me$ is involved.

The first source of subjectivity of $E(L_R,t)$ if the location $L$ where it occurs, with usually  $L\subseteq L_R$, and its time interval $\Delta T$, with usually  $\Delta T \subseteq \Delta T_R$. Thus, for instance, if $L_R$ is a city, then $me$ may be driving in some street or may be eating at home. Being in a location at a certain time, $me$ does not have access to what is going in the other locations and partially also in that location, if big enough. 
We capture this request by refining Eq. (\ref{eq:C_RE1}) into Eq. (\ref{eq:C_RE2}) below.

\vspace{-0.4cm}
\begin{equation}
C_P =  \langle  L_R, \{ E(L, t)\} \rangle \ \ \ \text{with}\ \ \   t\in \Delta T\subseteq\Delta T_R,\ L\subseteq L_R 
\label{eq:C_RE2}
\end{equation}

\vspace{-0.2cm}
\noindent
The second and most important cause of the subjectivity of events involving $me$ is that, as from above: (i) events are the result of the interactions among entities; (ii) these interactions happen because of the actions of entities; and (iii) these actions are motivated by the entities' mutual functions. Thus, for instance, $me_1$ can be in the \textit{car} of \textit{her friend} $me_2$, \textit{ in front} of the \textit{church}, while \textit{talking} to a \textit{friend} $me_3$.
We model this form of subjectivity by defining $E(L,t)$ as follows. 

\vspace{-0.1cm}
\begin{equation}
E(L,t)= \langle L, \{ \langle e, F_e, A_{F_e}(t) \rangle\}\rangle\ \text{with}\  t\in \Delta T
\label{eq:C_P}
\end{equation}

 \noindent 
where: $\Delta T\subseteq \Delta T_R$, $ L\subseteq L_R $, and where

\vspace{-0.2cm}
\begin{itemize}
    \item 
$e\in \{ e \}_{E}$ is any \textit{entity},  including $me$, involved in $E(L,t)$, with
$\{ e \}_{E}\subseteq \{e (O)\}_S$ the set of such entities, $\{ e \}_{E} \cup \{e\}_R \not = \emptyset$; 

\item $F_e=\{ f \}_e$ is the set of functions $f$ of any $e\in \{ e \}_{E}$ with respect to any $e\in \{ e \}_{E}$;

\item
$A_{F_e}= \{ a \}_{F_e}$ is the set of actions $a$ of any entity $e\in\{ e \}_{E}$ towards any other $e\in \{ e \}_{E}$, because of a function $f\in F_e$.
\end{itemize}
\vspace{-0.2cm}
\noindent
We can now merge Eq. \ref{eq:C_RE2} and Eq. \ref{eq:C_P} to obtain  the following final characterization of the subjective context $C_P$ of $me$:

\vspace{-0.2cm}
\begin{equation}
C_P =  \langle  L_R, \{
\langle L, \{ \langle e, F_e, A_{F_e}(t) \rangle\}\rangle\
\} \rangle\
\label{eq:C_P2}
%\vspace{-0.1cm}
\end{equation}
with  $t\in \Delta T \subseteq \Delta T_R, L\subseteq L_R$.
That is, looking at the pairs $\langle ... \rangle$ in Fig. \ref{eq:C_P2}, the subjective context of $me$ is constructed as follows:

\vspace{-0.2cm}
\begin{enumerate}
\item select a previously identified spatial region $S$, for instance the home of $me$, a museum, the university, or a city;
\item select a previously identified reference context $C_R$, that is, observation period $\Delta T_R$, reference location $L_R$ and entities $\{e\}_R$;
\item then select a set of $me$'s, each with a corresponding personal context $C_P(me,C_R)$; and,
\item for each $C_P$ select a set of events, each event with its own set of entities and corresponding, objects, functions and actions.
\end{enumerate}
\vspace{-0.2cm}
\noindent
where the four steps above are performed based on a given purpose.
Some observations. The first is about how \textit{subjectivity} is modeled. As from above, usually $me$'s do not know about the other $me$'s, this because of their different locations and time periods.
But this applies also to the entities involved in the same event. In fact, the functions and also the actions relating people - and entities in general - to one another, are unknown to most $me$'s. Subjectivity arises because of the \textit{diversity of people} and because of the partial knowledge that any $me$ has of the other $me$'s. This is the key intuition behind the idea of big-thick data. The only way to know about context is to ask people. Big data cannot provide information that thick data provide. 

The second observation is that we have assumed that, inside events, locations, functions and therefore entities are time-invariant. Their time variance, has to be managed by splitting an event in two or more, following a process similar to that of time-variant $C_R$'s.

The third observation is about \textit{locations}. The same spatial region $R$ can play the function of entity or location. Thus for instance, if $me$ is driving home, then home is the entity that $me$ needs to reach. It is outside the space occupied by $me$ and both $me$ and home are inside the same location, e.g., the city of Trento. However, when $me$ is at home, home is the smallest location outside $me$. Similarly, when $me$ is in the car, the car is the location where the driving event occurs but it is also the entity taking $me$ home. Any object can act as location: a person's body is the location where COVID-19 operates, a piece of sheet is the location where $me$ is writing, and so on. The discriminating factor is the space granularity of the event. 

The fourth observation is about \textit{entities} $e\in\{ e \}_{E}$ and their functions. Any $e$ may be concurrently involved in multiple events, usually from different $me$'s, usually with different functions. The functions of the entities  $e\in\{ e \}_{R}$ are known by all $me$'s, those of the entities  $e\not\in\{ e \}_{R}$, instead, are known only to some $me$'s. Thus, for instance, if $me_1$ meets $me_2$ at the University, $me_1$ will know that the University is her own study place and will not know that it is the work place for $me_2$, and vice versa. 

\begin{figure}[h]
\centering
\includegraphics[width=0.47\textwidth]{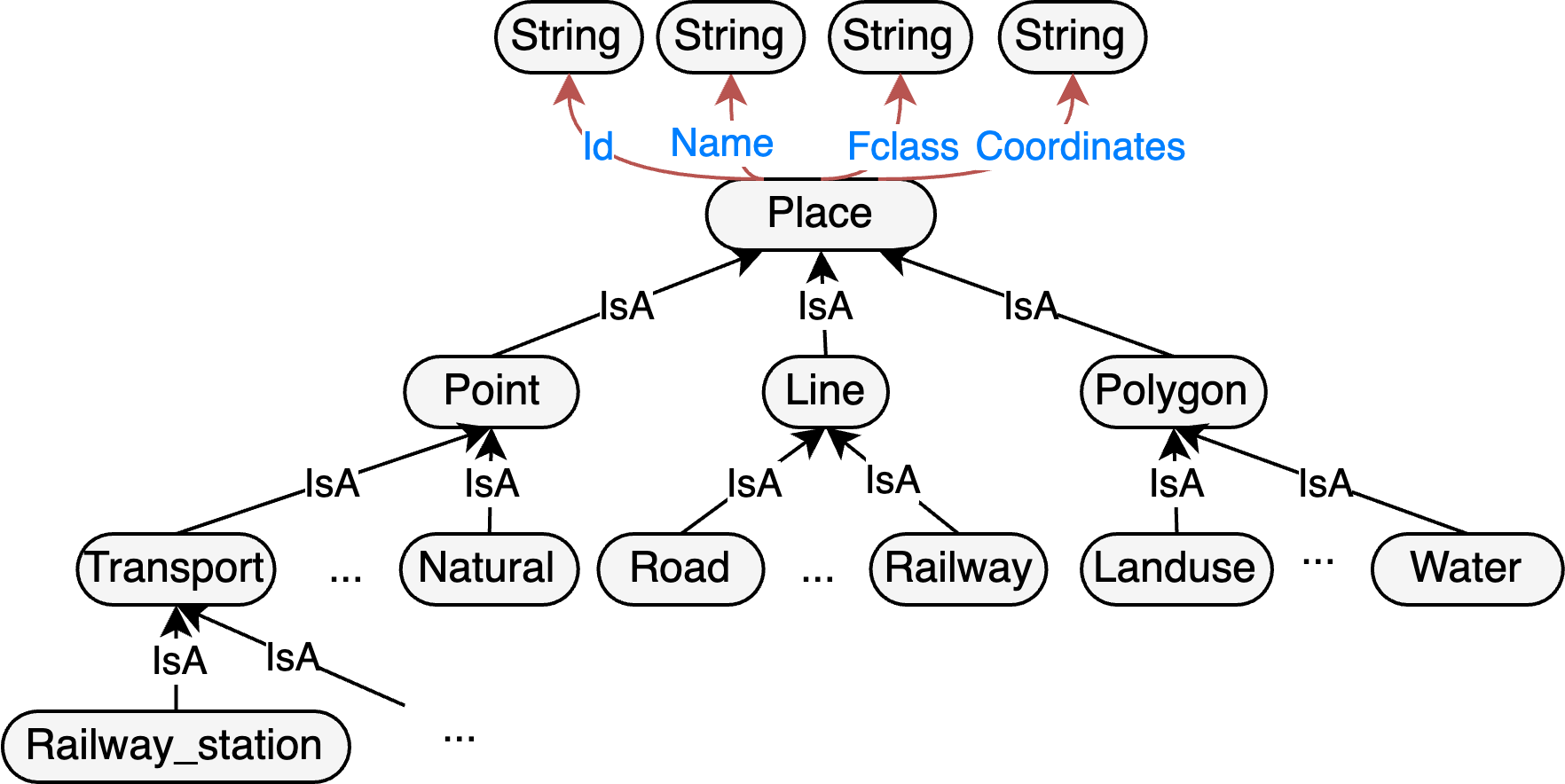}
\caption{The OpenStreetMap Hierarchy.} 
\label{fig:OSM}
\vspace{0.4cm}
\end{figure}

\noindent
The fifth observation is about \textit{actions} $\{ a \}_{F_e}$. Differently from functions, actions change frequently in time, with duration $\Delta t\subseteq \Delta T$, and this is why Human Activity Recognition (HAR) is hard. Big-thick data provide extra information, in particular, with respect to the functions that actions are supposed to carry out, see, e.g., \cite{KD-2021-Zhang-putting}.

Last but not least, so far we have talked of the functions and actions which characterize how an entity interacts with the outside. However, the behaviour of entities, and humans in particular, is largely influenced by their internals.
This is why big-thick data often carry information about people's \textit{internal functions and characteristics}, e.g., personality and procrastination syndrome, that we assume stable in time, and \textit{internal actions}, e.g., mood and tiredness, with usually relevant temporal dynamics \cite{KD-2022-Bontempelli-lifelong}.

%% file: section/4-unification.tex
Following \cite{giunchiglia2017personal}, we represent contexts as Knowledge Graphs (KGs) \cite{hogan2021knowledge,ji2021survey}. 
Let us consider $C_R$, see Eq. (\ref{eq:C_R2}).
We exemplify the process of building $C_R$ by formalizing the fragment of the OSM hierarchy in Fig. \ref{fig:OSM}.\footnote{Fig. \ref{fig:OSM} is depicted with reference to the OSM Layered GIS Format, see \url{https://download.geofabrik.de/osm-data-in-gis-formats-free.pdf}{\label{OSM Layered GIS Format}}.}
In Fig. \ref{fig:OSM}, the upper part of the hierarchy, down to the level of \texttt{Point}, \texttt{Line} and \texttt{Polygon}, describes the three types of geometrical features of \texttt{place}s.  
Each such feature, in turn, is further refined into a sub-hierarchy of max depth 4, as in 
\texttt{Place < Line < road < major\_road < motorway}.
The attributes of \texttt{Place} must be interpreted as follows: \texttt{Id} is its identifier; \texttt{Name} is its local name; \texttt{Fclass} 
is the property used to name its class; \texttt{Coordinates} is used to store its spatial coordinates. 

\begin{figure}[ht]
\vspace{-0.5cm}
\centering
\includegraphics[width=0.45\textwidth]{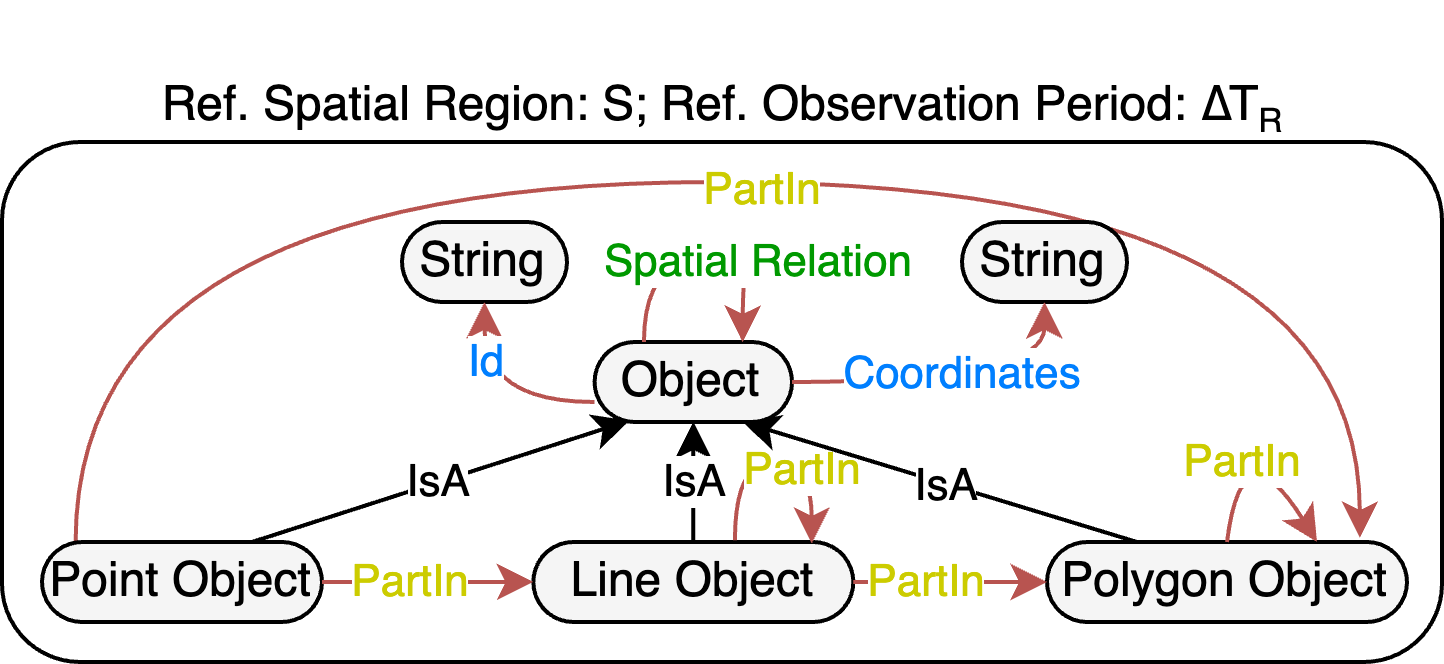}
\vspace{0.1cm}
\caption{The $C_S$ Teleontology.} 
\label{fig:STLO}
\vspace{0.5cm}
\end{figure}
\noindent
We proceed in steps, as follows. If one looks at Fig. \ref{fig:OSM}, it should come intuitive that the upper part encodes the geometrical features of objects and, as such, it should be represented in $C_S$, while the rest describes the properties, mainly functions, of entities and, as such, should be represented in $C_R$. Let us start from $C_S$. 
The first step is to produce the \textit{Space (Context) Teleontology (STLO)} \footnote{Following \cite{KD-2017-ICCMF}, the meaning of the word \textit{teleontology} builds on the Greek words \textit{telos} (meaning \textit{end, purpose}) and \textit{logia}, (meaning \textit{a branch of learning}). 
We use the word teleontology (and teleology, see below) to capture the intuition that a a teleontology is written with a \textit{purpose}. There is no claim of generality beyond the purpose for which it is generated.}, as represented in Fig. \ref{fig:STLO}. STLO  is a KG, representing a generic $C_S$, as from Eq. (\ref{eq:C_R1}), where nodes are \textit{object types} (with the added  \textit{datatype} \textit{String}), i.e., sets of objects associated with a set of properties. Nodes are connected by three types of links: (i) the subsumption relation \textit{IsA} rooted in the type \textit{Object}  denoting all objects;
(ii) \textit{Object Properties} linking two object types, i.e., \textit{Spatial Relation} and \textit{PartIn}, and (iii) two \textit{Data Properties}, i.e., \textit{Id}, \textit{Coordinates}.
STLO has four  main purposes: (i) to represent all the objects of $C_S$; (ii) to represent their selected spatial functionality  i.e., \textit{Point (Object)}, \textit{Line (Object)}
and \textit{Polygon (Object)}; 
and (iii) to represent spatial containment, i.e. \textit{PartIn},  as from Eq. (\ref{eq:L}), and (iv) spatial relations, see Section \ref{sec:Defining Context}.
\textit{PartIn} is applied recursively, thus allowing to define locations, as polygons or lines, at any level of spatial granularity, where locations, at the end of the recursion, are populated by point objects. The property \textit{Spatial Relation} can be specialized to more refined relations, e.g., \textit{Near}, \textit{Right} or \textit{North}. Both \textit{PartIn} and \textit{Spatial Relation} can be computed from OSM.
Finally, the box around the Teleontology KG represents the region $S$ and observation period $\Delta T_R$  of $C_R$. Implementationally, the information about boxes (all of them, see also below) is encoded as metadata.
 
\begin{figure}[h]
\vspace{-0.5cm}
\centering
\includegraphics[width=0.45\textwidth]{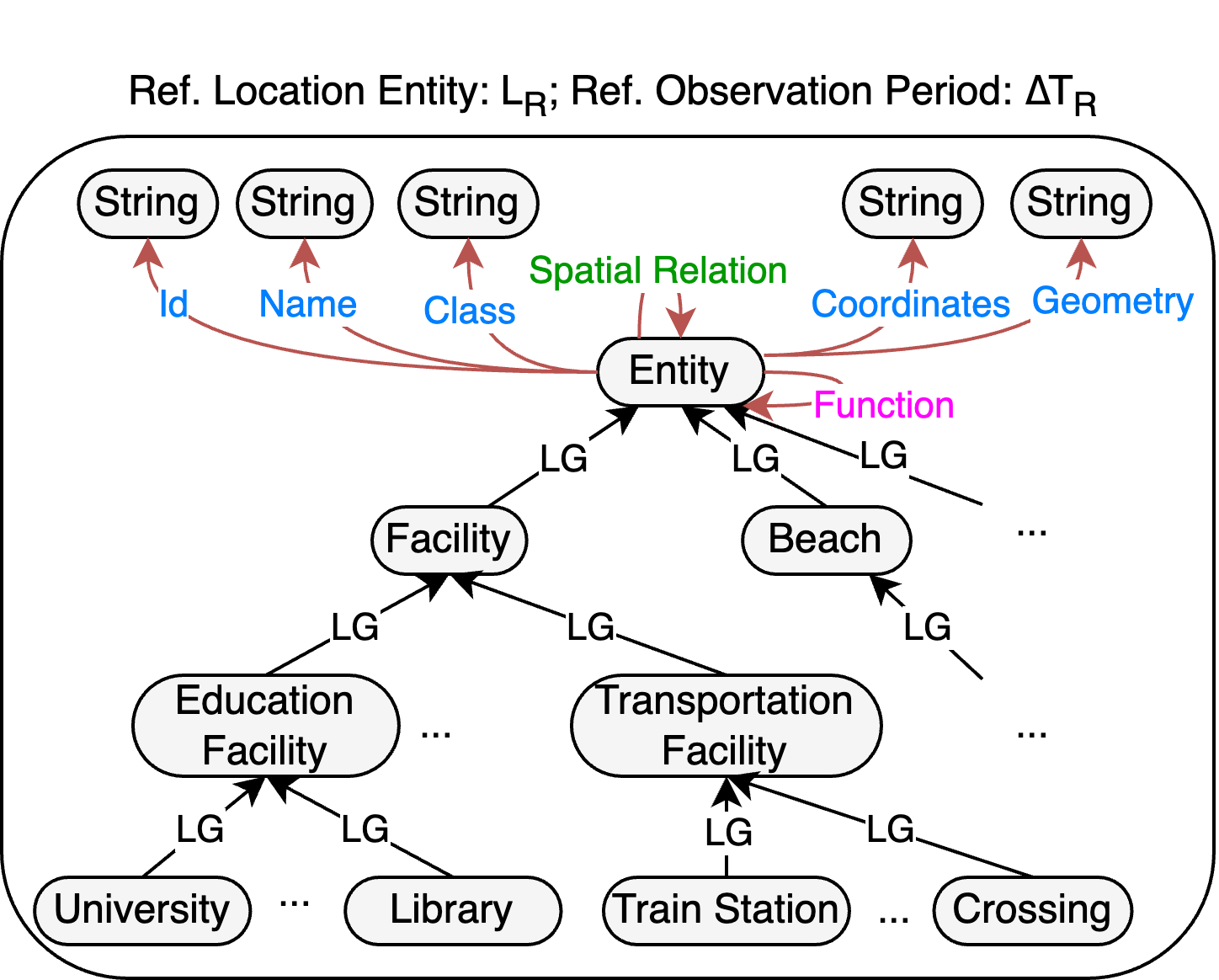}
\vspace{0.1cm}
\caption{The $C_R$ Teleontology.} 
\label{fig:KTLG}
\vspace{0.5cm}
\end{figure}

\vspace{-0.1cm}
\noindent
Fig. \ref{fig:KTLG} depicts the \textit{Knowledge (Context) Teleontology (KTLO)} representing a generic $C_R$ as derived from  STLO in Fig. \ref{fig:STLO}. KTLO is a ‘standard' \textit{More/Less General (MG/LG) hierarchy} where nodes are \textit{entity types (etypes)} with root the etype \textit{Entity} and where the lower etypes inherit the properties of the higher etypes. 
There are three observations. The first is that the the root of KTLO and, therefore each entity, inherits the spatial properties of the root of STLO, where the data property \textit{Geometry} defines whether it is a point, a line or a region. These properties are additional with respect to the ones defining the function(s) of entities, e.g., the type of facility, see the property \textit{Function} in Fig. \ref{fig:KTLG}. The second observation is that a point object may be modeled as an entity, a polygon object as a location (i.e., as the label of the surrounding box). Thus $L_R$, the label of the box in Fig. \ref{fig:KTLG}, can be, e.g., \textit{home}, \textit{street} or \textit{city}. An entity with both types of geometry can act as either an entity or a location. The third is that entities are associated a \textit{Name}, and a \textit{Class}. We name entities and not regions, and the same region may take two different names if it changes its function (see above the \textit{chicken coop} example). \textit{Class} is our term for \texttt{Fclass} in OSM.  The \textit{Class} 
of an etype can take as value the etype itself or more specific one, for instance we want to say that, e.g., a specific \textit{Facility} is a \textit{Pub}.

\begin{figure*}[ht]
\centering
\includegraphics[width=0.98\textwidth]{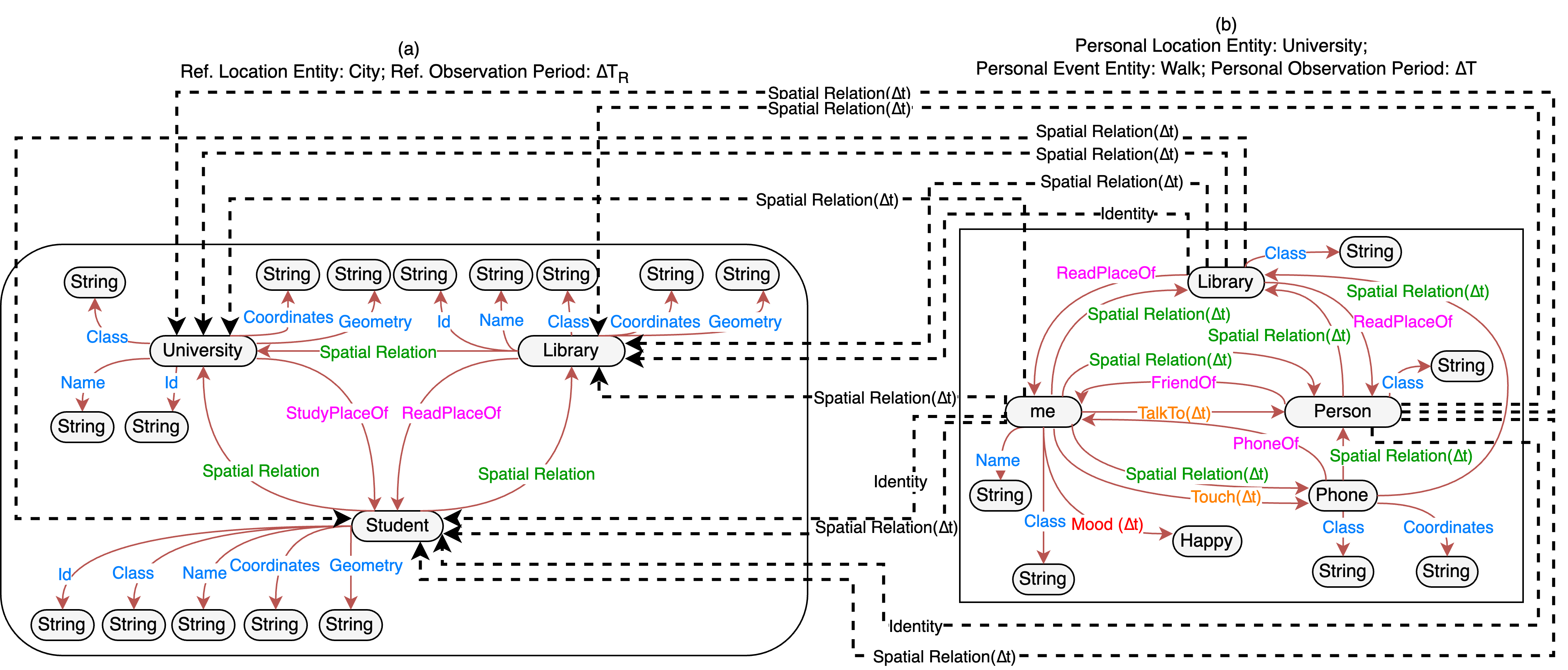}
\vspace{0.1cm}
\caption{(a) $C_R$ Teleology, and  (b) $C_P$ Event Teleology, unified.
}
\vspace{0.3cm}
\label{fig:Etyperelations}
\end{figure*}

Finally, we can now represent $C_R$ as an \textit{Entity Type (eType) Graph (ETG)} \cite{giunchiglia2021stratified}, that we also call a \textit{Teleology}. 
ETGs are obtained from KTLO's as follows: (i) select a subset of the etypes of the KTLO; (ii) for each etype, select a subset of its object and data properties, and, finally (iii) eliminate the \textit{LG} relation in KTLO and, for each type, distribute its properties to the lower etypes.  The generation of an ETG from a KTLO is the process of selecting what is relevant to the specific purpose, where, here, the purpose is to build an ETG which allows for context unification with the $C_P$ of one or more $me$'s. A fragment of the ETG resulting from Fig. \ref{fig:KTLG} is reported in Fig. \ref{fig:Etyperelations}(a). Fig. \ref{fig:Etyperelations} reports the ETG of $C_R$ (Fig. \ref{fig:Etyperelations}(a) on the left) together with the ETG of one $C_P$ (Fig. \ref{fig:Etyperelations}(b) on the right) which need to be unified. 

The $C_P$ ETGs are constructed following the same process, described above for the $C_R$ ETG (from STLO to KTLO to ETG), with a few key differences that we describe below.
Let us compare  $C_R$ and $C_P$ ETGs in Fig. \ref{fig:Etyperelations}. The first observation is that $C_P$ and $C_R$ share the same location entity \textit{City}, this because we have assumed that the event does not occur in a sub-location $L\subseteq L_R$. Still we have $\Delta T \subseteq \Delta T_R$. A complete visualization of $C_P$ could have been that $me$ was first reading in the \textit{Library}, then walking towards the \textit{University}, and then taking a class inside the \textit{University}. $C_R$ and $C_P$ share some entity types, e.g., \textit{Library}, namely those entities which, inside $C_R$ are relevant to the activities of $me$. They also share \textit{Spatial Relation}s, and \textit{Coordinates}, which is what allows to position $me$ inside $C_R$ while she moves around. The key difference is that the spatial relations involving $me$ are time-dependent. Both $C_R$ and $C_P$ involve functions, some of which, e.g., \textit{StudyPlaceOf},  define the function of an entity in $C_R$ with respect to $me$. Notice how functions are triples <entity, function, entity>. Differently from $C_R$, $C_P$ contains also actions, both external, e.g., \textit{TalkTo} and internal, e.g., \textit{Mood}. We also model actions as triples but, differently from functions, actions have a time duration $\Delta t \subseteq \Delta T$ that sometimes reduce to timestamps. 

Given the $n+1$ EGs the final step is to populate them with the available data,  for instance using the approach described in \cite{giunchiglia2021stratified,KD-2022-Bocca}. In the case study in Section \ref{sec:Case Study}, $C_R$ is populated with OSM data about \textit{Trentino}, the region where Trento is located, the $C_P$'s with SU2 data. The result is one $C_R$ \textit{Entity Graph (EG)} and $n$ $C_P$ EGs, one each each $me$. EGs are KGs where nodes are specific entities, e.g., the buildings of Trento or a specific person associated to a $me$, each associated with its own etype, and links are the properties of the corresponding ETG. Intuitively, EGs are built from ETGs by expanding each and every etype node into all the entities of that etype, e.g., \textit{Restaurant(Biba's)}, \textit{Library(BUC)}, 
 and \textit{me(User73)}, and by adding one link for each specific instantiation of a property. Some observations. The first is that the entities can be associated with spatial relations, e.g., \textit{Near(UniTn, BUC)} in  $C_R$  or \textit{Near(User73, User45, $\Delta t$)} in the \textit{User73}'s EG.
 The second observation is that, as a consequence of the fact that actions are tagged with a timestamp, for each $me$ we have a timed sequence of $C_P$ EGs, one for each selected duration period. That is, for each $me$, we have a \textit{stream of Spatio-Temporal} EGs.

Given one reference context and $n$ personal contexts, the next step is to \textit{context-unify} them and build the \textit{Observation Context} $C$.
Let us assume that the EGs of $C_R$ and the $C_P$'s, as represented in Fig. \ref{fig:Etyperelations}, have been constructed, Then, we have

\vspace{-0.3cm}
\begin{equation}
    C\ =\ C_R\ \uplus\ \{ C_P\}
\end{equation}

\vspace{-0.2cm}
\noindent
where $\uplus$ is the \textit{context unification operator} and $ \{ C_P\}$ is the set, one or more, of personal contexts under consideration.  Context unification operates in two macro steps, as follows.

\vspace{-0.1cm}
\begin{itemize}
    \item Unification between $C_R$ and $\{C_P\}$, one $C_P$ at the time;
    \item Pairwise unification between any two $C_P$'s after they are unified with $C_R$.
\end{itemize}

\vspace{-0.2cm}
\noindent
The order of unification is motivated by the fact that the first step objectifies, for what is possible, the contents of $C_P$ with respect to their position and also the functions relating the entities in $C_R$ with the entities in a $C_P$. This allows to obtain new properties, such as, e.g., \textit{Near(User73, Biba's,  $\Delta t$)}, \textit{RestaurantOf(User73, Biba's,  $\Delta t$)}. Notice that we perform the unification of two personal contexts always with respect the reference context.

Context unification exploits three specific types of unification, as follows.

\vspace{-0.2cm}
\begin{itemize}
\item \textit{Etype} and \textit{Property Unification (EPU)}. This is typical problem of \textit{ontology / schema alignment}; we follow the approach described in \cite{giunchiglia2020entity,shi2024kae};

 \item \textit{Spatio-Temporal Unification (STU)}. Here the spatio-temporal coordinates of entities are exploited. We have two types of results. The first is the recognition of two entities, for instance two $me$ belonging to two different $C_P$'s, as being the same entity. For this to be the case, the coordinates of the two entities must be the same, modulo approximations, at all times. The second is the spatial relation holding at a certain time between entities, see the examples above;

 \item \textit{Entity Unification (EU)}. This is done using specific entity properties, different from spatio-temporal properties, mainly \textit{Name} and \textit{identifier}, but also entity specific attributes \cite{giunchiglia2021stratified}. For instance, if available, information about the phone number. 
 
\end{itemize}

\vspace{-0.1cm}
\noindent
Fig. \ref{fig:Etyperelations} shows the possible unifications between $C_R$ (Fig. \ref{fig:Etyperelations}(a) and one $C_P$ in Fig. \ref{fig:Etyperelations}(b). For instance, we could learn that \textit{User73}, the specific $me$ in Fig. \ref{fig:Etyperelations}(b) is in \textit{Trento} in certain period, and near \textit{BUC} at a certain hour, that he is talking to \textit{User72} half an hour later, that the two are friends, and so on.

%% file: section/5-observation.tex
The context unification process leaves full flexibility in the selection of the reference context, of the personal contexts, and also what to unify. 
The result is a spatio-temporal EG which can then be \textit{enquired} in many different ways, for instance, it can be queried like any other KG \cite{KD-2022-Giunchiglia.a}, it can be used to do statistical modelling and reasoning \cite{bison2024impacts}, or it can be used to do machine learning, for instance for the machine to learn about the user and thus to enable high quality high value human-machine interactions \cite{KD-2020-Bontempelli1}. 
Because of this, the approach proposed in this paper can be seen as defining a general purpose \textit{meta-process} which can be used to build big-thick data for the desired purpose. The case study in Section \ref{sec:Case Study} provides a relatively large example of how big-thick data can be generated and then used to do prediction. In this perspective it becomes relevant to classify the possible \textit{observation purposes} into four main groups, as a function of what one is interested in observing. We have the following.  

\vspace{-0.2cm}
\begin{itemize}
    \item \textit{Reference (R) enquiries}. The goal here is to know the details of the reference context, for instance as someone would do when getting to a new place for the first time and being in need of finding her way around.
    R-enquiries are posed \textit{only} to the reference context, independently of the dynamics which may occur inside it. Thus, for instance, possible questions which could be posed to the observation context built in Section \ref{sec:Case Study} are:
    ‘Where is the bus stop near to the bar named Bar Sport?' or ‘What are the supermarkets near BiBa's?'.
    
    \item \textit{Personal (P) enquiries}. The goal here is to know about what people have done in certain period of time, including also their subjective view of what happened. P-enquiries are only to a single $C_P$ (or streams of $C_P$'s of the same $me$, see below) with no possibility of reference to entities of $C_R$ which are not part of $C_P$. For instance, in the case study in Section \ref{sec:Case Study}, possible questions which could be posed are ‘Which places did $me$ go in a certain period, what she did there, and what was her mood?'. The answers are associated with \textit{me}'s subjective locations (e.g., shopping place), events (e.g., shopping), social interactions (e.g., a seller) and moods (e.g., happy). 

      \item \textit{Personal-Reference (PR) enquiries}. The goal here is to explore how the inside of the reference context evolves as a function of the entities which populate it within a certain period of time. PR-enquiries are posed to the observation context $C$ focusing on $C_R$ and are about its state as a function of the activities of one or more $C_P$'s. Some examples from the case study are: ‘How many people are in the Biba's restaurant during week-ends?' and ‘How many attractions in Trento have involved $me$ during the last week?’.
      
    \item \textit{Reference-Personal (RP) enquiries}. The goal here is to explore the environment around $me$'s and its impact on $me$. These are enquiries about one or more $me$ posed to the observation context $C$. Some examples are ‘What was your mood when you were in the Coop supermarket?' and ‘which friends of $me$ were in the Biba’s restaurant during the dinner, yesterday?'.
\end{itemize}

\vspace{-0.1cm}
\noindent
Roughly speaking, R-enquiries correspond to the ‘typical' problems dealt with using big data, while P-enquiries relate to the ‘typical' problems dealt with using thick data. PR-enquiries are new types of enquiries, enabled by big-thick data, where one can study how specific elements of objective knowledge can be enriched by the subjective knowledge provided by multiple people, as a function of their behaviour and subjective perspective. Similarly, RP-enquiries are new types of queries, enabled by big-thick data, which allow us to put the subjective behaviour of people under the objective lenses of a third party, possibly also providing an cross-individual inter-subjective view of the world.

%% file: section/6-CaseStudy.tex
We generate big-thick data by unifying OSM big data with the SU2 thick data. We first build the $C_R$ EG, then the $C_P$ EGs, then the observation context EG, which we then enquire.

\subsection{The Reference Context EG}
\label{sec-csrc}

The Geofabrik site\footnote{\url{https://download.geofabrik.de}{\label{Geofabrik}}} provides geographical information worldwide about physical places, that we call OSM \texttt{Place}s, each associated with multiple features and classes. OSM \texttt{Place}s are associated with various data properties, some shared by all \texttt{Place}s, e.g., \texttt{id}, \texttt{Name}, \texttt{Fclass} and \texttt{Coordinates}, some others associated only to specific \texttt{Place}s, e.g., \texttt{type} is a property of \texttt{Building}, with values, e.g., \texttt{apartment} or \texttt{church}. We construct the \textit{OSM-Trentino} dataset, in SHP format, by constraining the places' coordinates to lie within the Trentino maximal and minimal latitudes and longitudes. 

\begin{figure}[h]
\vspace{-0.3cm}
\centering
\includegraphics[width=0.47\textwidth]{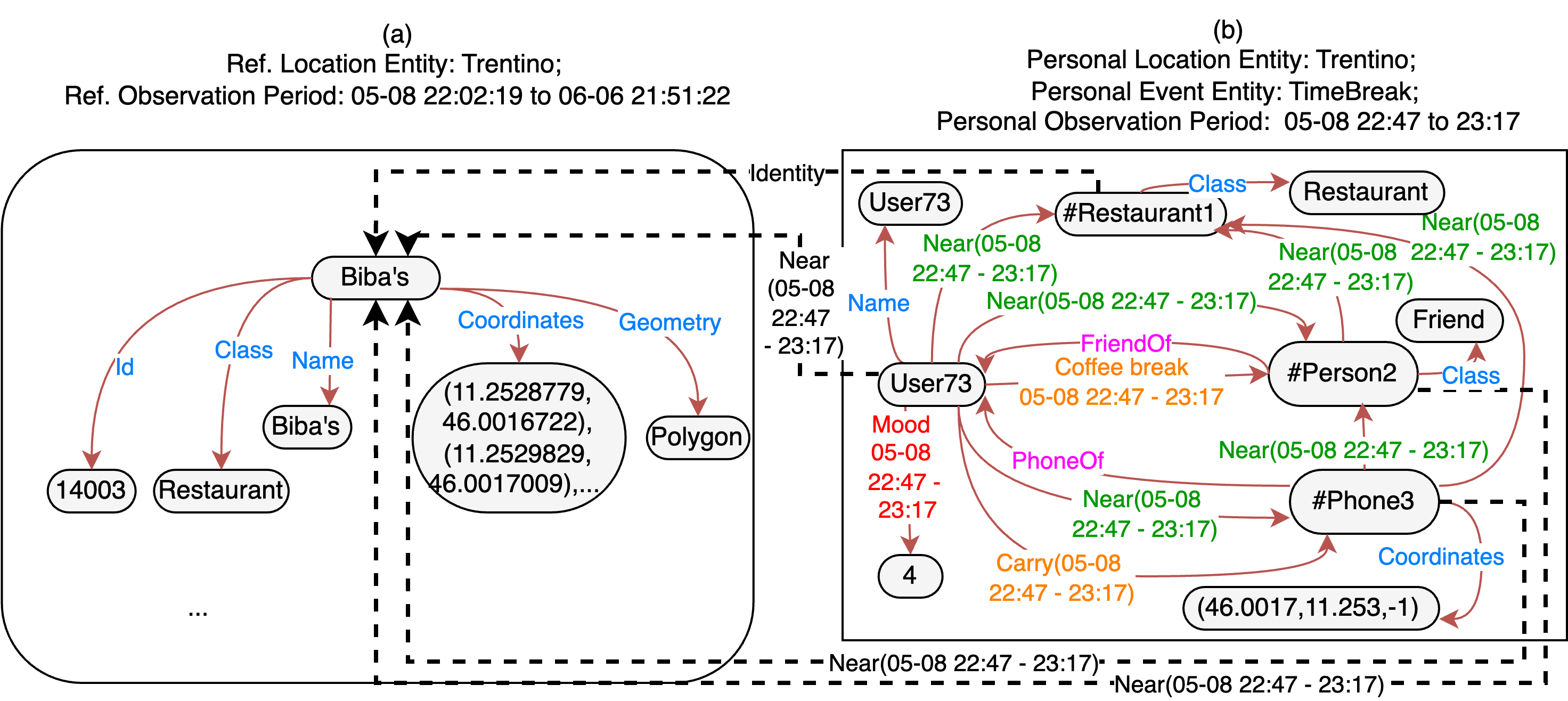}
\caption{An Observation Context EG from the SU2OSM Dataset.} 
\label{fig:UEGExample}
\end{figure}

We construct the $C_R$ from the \textit{OSM-Trentino} dataset using the properties mentioned in the previous sentence. The $C_R$ reference observation period $\Delta T_R$  is the period of the SU2 data collection experiment, that is, four weeks from 05-08 22:02:19 to 06-06 21:51:22, where the year is removed for privacy reasons.
 We define place entities, one for each OSM \texttt{Place}, and we compute spatial relations among them, e.g., \textit{Near}, from their coordinates.  For instance, a fragment of the \textit{Trentino} $C_R$ EG which described the restaurant of name \textit{Biba's} is represented in Fig. \ref{fig:UEGExample}(a). 

\vspace{-0.1cm}
\subsection{The Personal Context EGs}
\label{sec-cspc}

The SU2 dataset\footnote{For detailed information about the \textit{SU2} dataset and the process by which has been generated, look at the link reported in footnote$^{\ref{foot:SU2}}$.} consists of phone sensor data, time diaries and surveys collected, during a period of four weeks,
from $158$ students of the University of Trento, using the iLog APP \cite{zeni2014multi,zhao2024human} running on their phones. 
The SU2 dataset is suitably anonymized, it abides by the General Data Protection Regulation (GDPR), and has been utilized in numerous case studies, see, e.g., \cite{KD-2019-Zeni-UBICOMP,KD-2021-Zhang-putting}.
The time diaries used in the data collection include various HETUS questions\footnote{More information about the \textit{HETUS (Harmonized European Time Use Surveys} standard can be found at \url{https://ec.europa.eu/eurostat/web/time-use-surveys.}}%, which were asked every thirty minutes during the first two weeks and every hour during the last two weeks
. The questions used in this case study are: \textit{Where}: `Where are you?'; \textit{What}: `What are you doing?'; \textit{WithWhom}: `With whom are you?'; plus, additionally, one question on \textit{Mood}: `What is your mood?'. The SU2 dataset also includes a wide range of sensor data. Here we use only the GPS. Plus we do not use the information collected during the surveys, before and after the data collection. To summarise, in this case study we use the following data from the SU2 dataset:

\vspace{-0.2cm}
\begin{itemize}
\item \textit{userid}: the participant identifier, valuing integers from 0 to 157 (the data of some students are considered because the quality was too low);
\item \textit{where}: the answer of the \texttt{Where} question, a total of 17 possible answers, including `Home', `Classroom / Study hall', etc;

\item \textit{what}: the answer of the \texttt{What} question, a total of 23 possible answers, including `Sleeping', `Eating' and `Studying', etc;

\item \textit{withWhom}: the answer of the \texttt{WithWhom} question, a total of 9 possible answers, including `Friend(s)', `Classmate(s)', etc;

\item \textit{mood}: the answer of the \texttt{Mood} question, valuing integers from 1 to 5. Higher values mean more positive mood;
\item GPS coordinates, i.e., \textit{latitude}, \textit{longitude}, and \textit{altitude};
\end{itemize}

\vspace{-0.2cm}
\noindent
GPS data, questions and respective answers are time-stamped, where one timestamp has form `mm-dd hh:mm:ss'. The frequency of questions is every 1/2 hour or every hour depending on the week. The GPS data has an estimated frequency of once a minute.

We construct the $C_P$'s as follows. We have one $me$ for each participant. 
This data is formalized into the $C_P$ Event Teleology in Fig. \ref{fig:Etyperelations}(b) in the obvious way, but with a few twists which take into account the specificity of the SU2 dataset. 
During the overall four weeks of the data collection there were a range of $1063 - 1067$ question batteries per user, $168,095$ in total, each battery involving the questions listed above (and more). We associate to each such battery a $C_P$ associated to a single event $E(L)$ of duration $\Delta T$, the same for both $C_P$ and $E(L)$. We assume that $\Delta T$  is the interval time between two questions, viz., half an hour and one hour in the first and second two weeks, respectively, centered in the time of the question. 
The result is a total of $104,414$ $C_P$'s (participants did not always answer).
organized into $158$ \textit{$C_P$ streams}, one for each $me$, with length in the range of $371-875$ EGs.
We call any element of the stream a \textit{timed EG (or $C_P$)}. 

To populate $C_P$'s,
one point of attention relates to the answers of the questions \textit{What}, \textit{Where}, and \textit{WithWhom}. The answers to the first question can be directly encoded as actions. However, the encoding of the answers of the last two questions is a little more elaborated. In fact, because of its intended use (mainly within the Social Sciences), HETUS is designed to allow for generic answers, what in the $C_P$ we encode as etypes. Thus, for instance, the $me$ of name \textit{User73} might answer that she is with a friend. Which means that $me$ is with a person entity, whose name is unknown. Thus, we translate the above two answers in the following triples: \textit{FriendOf( \#Person2, User73)}, \textit{WithWhom(User73, \#Person2, $\Delta t$)}, \textit{Near(User73, \#Person2, $\Delta t$)}, \textit{Near(User73, \#Restaurant1, $\Delta t$)}, where these anonymous entities have the obvious etypes. The time parameter $\Delta t$ encodes when that time-variant spatio-temporal property holds. Figure \ref{fig:UEGExample} 
(b) reports a fragment of a \textit{TimeBreak} event involving  the participant \textit{User73}. 
A second point of attention relates to the spatio-temporal position of entities.
We computed the coordinates of phones by considering the coordinates in a time window of $10$ minutes around the question time. Then, we apply DBSCAN \cite{schubert2017dbscan} and identify the largest cluster computed by the algorithm. The coordinates are the mean of the coordinates of this cluster. 

\begin{table*}[htp!]
\centering
\caption{Prediction enquiries.}
\scalebox{0.9}{
\begin{tabular}{|c|c|c|ccccccc|}
\hline
\multirow{2}{*}{\textbf{\begin{tabular}[c]{@{}c@{}}Prediction \\ Enquiry\end{tabular}}}    & \multirow{2}{*}{\textbf{Dataset}} & \multirow{2}{*}{\textbf{\begin{tabular}[c]{@{}c@{}}Purpose  \\ Feasibility\end{tabular}}} & \multicolumn{7}{c|}{\textbf{Prediction Experiment}} \\ \cline{4-10} 
 & &      & \multicolumn{1}{c|}{\textbf{\begin{tabular}[c]{@{}c@{}}Target \\ Property\end{tabular}}} & \multicolumn{1}{c|}{\textbf{Feature Properties}} & \multicolumn{1}{c|}{\textbf{\begin{tabular}[c]{@{}c@{}}Best-performance\\  Algorithm\end{tabular}}} & \multicolumn{1}{c|}{\textbf{Accuracy}} & \multicolumn{1}{c|}{\textbf{Recall}} & \multicolumn{1}{c|}{\textbf{F1 Score}} & \textbf{AUC} \\ \hline

 % Q1-1 row
\multirow{3}{*}{\begin{tabular}[c]{@{}c@{}}E1: Does a place \\ is classified as \\a residence?\end{tabular}} & OSM-Trentino  & $\checkmark$  & \multicolumn{1}{c|}{type} & \multicolumn{1}{c|}{name, class}   & \multicolumn{1}{c|}{Random Forest}  & \multicolumn{1}{c|}{78.23\%}    & \multicolumn{1}{c|}{0.529}     & \multicolumn{1}{c|}{0.491}    & 0.529     \\ \cline{2-10} 
% Q1-2 row
 & SU2    & $\times$   & \multicolumn{1}{c|}{-}    & \multicolumn{1}{c|}{-}& \multicolumn{1}{c|}{-} & \multicolumn{1}{c|}{-}   & \multicolumn{1}{c|}{-} & \multicolumn{1}{c|}{-}   & -     \\ \cline{2-10} 
% Q1-3 row
 & SU2OSM& $\checkmark$  & \multicolumn{1}{c|}{type} & \multicolumn{1}{c|}{\begin{tabular}[c]{@{}c@{}}day\_of\_week, time\_of\_day,\\  name, class\end{tabular}} & \multicolumn{1}{c|}{Random Forest}  & \multicolumn{1}{c|}{82.51\%}    & \multicolumn{1}{c|}{0.714}     & \multicolumn{1}{c|}{0.728}    & 0.825      \\ \hline

%Q2-1 row
\multirow{3}{*}{\begin{tabular}[c]{@{}c@{}}E2: Is a user \\ at a living place?\end{tabular}}    & OSM-Trentino  & $\times$   & \multicolumn{1}{c|}{-}    & \multicolumn{1}{c|}{-}& \multicolumn{1}{c|}{-} & \multicolumn{1}{c|}{-}   & \multicolumn{1}{c|}{-} & \multicolumn{1}{c|}{-}   & -     \\ \cline{2-10} 
 & SU2    & $\checkmark$  & \multicolumn{1}{c|}{\multirow{2}{*}{where}}     & \multicolumn{1}{c|}{what, withWhom, mood}  & \multicolumn{1}{c|}{Decision Tree}  & \multicolumn{1}{c|}{87.43\%}    & \multicolumn{1}{c|}{0.850}  & \multicolumn{1}{c|}{0.855}    & 0.930      \\ \cline{2-3} \cline{5-10} 
 & SU2OSM& $\checkmark$  & \multicolumn{1}{c|}{}     & \multicolumn{1}{c|}{\begin{tabular}[c]{@{}c@{}}name, class, what,\\ withWhom, mood\end{tabular}}   & \multicolumn{1}{c|}{Random Forest}  & \multicolumn{1}{c|}{94.42\%}    & \multicolumn{1}{c|}{0.846}  & \multicolumn{1}{c|}{0.878}    & 0.949     \\ \hline

 % E3
\multirow{3}{*}{\begin{tabular}[c]{@{}c@{}}E3: Is a user \\ in a bank?\end{tabular}}      & OSM-Trentino  & $\times$   & \multicolumn{1}{c|}{-}    & \multicolumn{1}{c|}{-}& \multicolumn{1}{c|}{-} & \multicolumn{1}{c|}{-}   & \multicolumn{1}{c|}{-} & \multicolumn{1}{c|}{-}   & -     \\ \cline{2-10} 
 & SU2    & $\times$   & \multicolumn{1}{c|}{-}    & \multicolumn{1}{c|}{-}& \multicolumn{1}{c|}{-} & \multicolumn{1}{c|}{-}   & \multicolumn{1}{c|}{-} & \multicolumn{1}{c|}{-}   & -     \\ \cline{2-10} 
 & SU2OSM& $\checkmark$  & \multicolumn{1}{c|}{class}    & \multicolumn{1}{c|}{what, where, withWhom, mood} & \multicolumn{1}{c|}{Decision Tree}  & \multicolumn{1}{c|}{90.91\%}    & \multicolumn{1}{c|}{0.731}  & \multicolumn{1}{c|}{0.736}    & 0.923      \\ \hline
\end{tabular}
}
\begin{tablenotes}
\footnotesize
\item Note 1. $\checkmark$ and $\times$ indicate whether E1, E2 and E3 succeed or fail, respectively, in satisfying the purpose feasibility metric of the selected dataset.

\item Note 2 (E1): \textit{target property} holds if the property \textit{type} (as from OSM)  has values ‘apartments' or ‘house' or ‘residential', it does not otherwise; the \textit{day\_of\_week} and \textit{time\_of\_day} features are labeled as the weekday (from Monday to Sunday) and the time periods (morning, afternoon, evening and night) based on the time of user answers.

\item Note 3 (E2): \textit{target property} holds if the property \textit{where} has values ‘Home', ‘Relatives Home' or ‘House (friends, others)', it does not otherwise.
\item Note 4 (E3): the \textit{target property} holds if the property \textit{class} has value ‘bank', it does not otherwise.
\end{tablenotes}
\label{tab:Es}
\end{table*}

\subsection{EG Unification}
\label{sec-csoc}

Given the contents of $C_R$, the purpose is to get to know about the physical places where SU2 events occur.
To achieve this, 
we unify the single OSM-Trentino EG with the streams of the SU2 EGs of the $158$ $me$'s into the observation EG, that we call SU2OSM. 
  Fig. \ref{fig:UEGExample} reports a fragment obtained from the union of the two EGs introduced before. The process proceeds in three steps as follows.

The goal of the \textit{first step} is to perform EPU unification (see Section \ref{sec:4-unification}). We have performed this step manually given that we are not interested in full automation. However the task is quite straightforward, given the limited scope of the etypes and properties of the $C_R$ and $C_P$ ETGs. Given the $C_R$ and $C_P$ KTLO's, this task is largely within the reach of the algorithm described in \cite{shi2024kae}.
The \textit{second step} is to establish, using STU unification, inside each timed $C_P$, the holding of spatial relations between $C_P$ phone entities and $C_R$ place entities. We focus on the spatial relation \textit{Near}. This is achieved, using their coordinates, by calculating the distances between phone entities and the closest OSM-Trentino place entities. We assume that two entities are one near the other if their distance is less than $50$ meters. For instance, in Fig. \ref{fig:UEGExample}, \textit{Near(\#Phone3, Biba's)} is the result of computing the distance of \textit{Biba's}, the  OSM-Trentino place closest to \textit{\#Phone3} as being approximately 8.2 meters. This allows us to derive which phone, place and person entities are close to one another. 
The \textit{third step}, performed using EU unification informed by STU unification, is to establish which specific $C_R$ place entity is the one where $me$ is. The idea is to select among the place entities which are close to $me$, the one with the proper etype. This allows us to identify a generic $C_P$ place entity, e.g., \textit{\#Restaurant1} in Fig. \ref{fig:UEGExample}, as being a specific $C_R$ entity, which in Fig. \ref{fig:UEGExample} is the restaurant of name \textit{Biba's}. 
In case of multiple places of the correct etype near the phone, we select the closest. Following the process described above, we have recognized a total of $147$ out of the $483,981$ OSM-Trentino entities, many of which matched more than once during the four weeks, thus highlighting the student \textit{habits}.
In total, $1955$ $C_P$ EGs have been unified with the $C_R$ EG (that is, $1.87\%$ of  the total number of timed $C_P$)
out of which we have computed $7820$ relations linking the $C_R$ EG to the $C_P$ EGs.
Notice how this would enable the unification of entities across $C_P$ EGs, allowing us, for instance, to establish when two different $me$ were in the same location at the same time, or how much time a single $me$ spent in the same place in certain periods (e.g., the morning of a specific time).

\vspace{0.1cm}
\begin{table}[]
\caption{OSM-Trento, SU2 and SU2OSM dataset size. }
\centering
\scalebox{0.75}{
\begin{tabular}{|c|c|c|c|c|}
\hline
\rowcolor[HTML]{EFEFEF}
\textbf{Dataset}      & \textbf{Storage Volume} & \textbf{CR EG} & \textbf{CP EG Streams} & \textbf{Unified EG Streams} \\ \hline
\textbf{OSM-Trentino} &           156,5 MB       & 1              & 0                      & 0                           \\ \hline
\textbf{SU2}          & 441,9MB                & 0              & 104,414                & 0                           \\ \hline
\textbf{SU2OSM}       & 19,2 MB                 & 1              & 104,414                & 1955                        \\ \hline
\end{tabular}
}
\label{tab:number}
\end{table}

\vspace{-0.2cm}
\noindent
Table \ref{tab:number} summarizes the dataset statistics. Notice how the SU2OSM dataset, while carrying more purpose relevant information than the union of $C_R$ and all the $C_P$'s has a size of around $19,2$MB, that is, around the $3\%$ of their cumulative total size ($598,4$MB), this resulting from the fact that a very small minority of the entities in the OS-Trentino dataset are relevant to the current purpose. This provides evidence of the scalability and effectiveness of the proposed approach in the generation of big-thick data. 

\vspace{-0.1cm}

\subsection{Observation Enquiries}
\label{sec-cs-oe}
We concentrate on three distinct binary classification prediction enquiries E1, E2 and E3, as from Table \ref{tab:Es}, which are illustrative examples of R-, P- and RP-enquiries, respectively (see Section \ref{sec:5-observ}). Each enquiry is identified by a \textit{target property} and a set of \textit{feature properties}, see Table \ref{tab:Es}. Target property and feature properties are the key elements characterizing the \textit{Purpose Feasibility} of a dataset (see column 3 in Table  \ref{tab:Es}). Purpose feasibility is a new boolean metric that indicates whether a dataset has the ability to support a certain purpose. 
We have tested various algorithms, i.e., \textit{Logistic Regression}, \textit{Decision Tree}, and \textit{Random Forest} following the \textit{5-fold cross-validation} approach \cite{kohavi1995study}. In the evaluation we have used standard metrics, as from Table \ref{tab:Es}, where AUC stands for \textit{Area Under the Curve}. 
 Table \ref{tab:Es} shows the results of prediction experiments for the E1, E2 and E3 with the best-performance algorithm. 
 
 There are two main observations.
The \textit{first} is about the purpose feasibility of SU2, OSM-Trentino and SU2OSM. OSM-Trentino can only answer E1 (a $R$ question) because it exclusively populates a $C_P$ EG, providing the target and properties for E1. 
    Dually, SU2 can only answer the E2 (a $P$ question), because it exclusively populates streams of $C_P$ EGs. 
    The merged SU2OSM dataset can answer all the proposed enquires because it unifies a $C_P$ EG with the streams of $C_P$ EGs. 
 The \textit{second} is about the prediction results. E1 can be answered both using OSM-Trentino and SU2OSM. However the SU2OSM metrics are better than those of OSM-Trentino, despite the first being much smaller. E2 can be answered both using SU2 and SU2OSM, but, again the SU2OSM metrics are better than those of the SU2 dataset. E3 can be answered only by SU2OSM.

%% file: section/7-Conclusion.tex
The main focus of the line of work described in this paper is the development of AI's which supports humans in their life in the real world, as distinct from the virtual world enabled by the Web. Within this application context, we propose using \textit{big-thick data}, namely, a new type of data which, in the opinion of the authors, are key and most likely necessary for the development of meaningful lifelong human-in-the-loop human-machine interactions.

The main result is an articulation of big-thick data as the result of the flexible integration, we call it \textit{context unification}, of reference context and personal context data.
The key element of this type of data is that it is not only machine generated, for instance in the form of IOT or Web data, but it is also \textit{purposely} provided by humans. It is \textit{only} humans who can provide high quality context-aware data. This human contribution can be in the form of the reference context, for instance as provided by motivated volunteers, as it is the case with OSM and open data, but it can also be in the form of personal data carrying detailed information of the \textit{why}, the \textit{what} and the \textit{how} of people's behaviour. Our future work will focus on how to tightly integrate big-thick data generation and machine learning.